# Integrating feature selection and regression methods with technical indicators for predicting Apple Inc. stock prices

Fatemeh Moodi[1,*], Amir Jahangard-Rafsanjani[2]

**Abstract**
Stock price prediction is influenced by a variety of factors, including technical indicators, which makes Feature selection crucial for identifying the most relevant predictors. This study examines the impact of feature selection on stock price prediction accuracy using technical indicators. A total of 123 technical indicators and 10 regression models were evaluated using 13 years of Apple Inc. data. The primary goal is to identify the best combination of indicators and models for improved forecasting. The results show that a 3-day time window provides the highest prediction accuracy. Model performance was assessed using five error-based metrics. Among the models, Linear Regression and Ridge Regression achieved the best overall performance, each with a Mean Squared Error (MSE) of 0.00025. Applying feature selection significantly improved model accuracy. For example, the Multi-layered Perceptron Regression using Forward Selection improved by 56.47% over its baseline version. Support Vector Regression improved by 67.42%, and Linear Regression showed a 76.7% improvement when combined with Forward Selection. Ridge Regression also demonstrated a 72.82% enhancement. Additionally, Decision Tree, K-Nearest Neighbor, and Random Forest models showed varying levels of improvement when used with Backward Selection. The most effective technical indicators for stock price prediction were found to be Squeeze_pro, Percentage Price Oscillator, Thermo, Decay, Archer On-Balance Volume, Bollinger Bands, Squeeze, and Ichimoku. Overall, the study highlights that combining selected technical indicators with appropriate regression models can significantly enhance the accuracy and efficiency of stock price predictions.

**Keywords** Stock price prediction; Technical indicators; Machine learning; Wrapper Feature selection methods; Regression methods; Time window.

## 1 Introduction

Predicting the stock market is a very complex task, and to accurately and efficiently forecast the market's future, various factors must be considered. Some of these factors include estimating future earnings, the publication of profit news, dividend announcements, management changes, and more [1]. A common method used by most traders to predict the stock market is technical indicators. This method helps analysts significantly reduce investment risks by providing accurate market analysis. Ignoring such analysis may lead to a loss of capital. Therefore, selecting the best combination of technical indicators alongside different machine learning methods for stock price prediction is of great importance.

[1] Department of Computer Engineering, Birjand University of Technology, Birjand, Iran.
[2] Department of Computer Engineering, Yazd University, Yazd, Iran.
* Corresponding author: moodi@birjandut.ac.ir

A technical indicator is essentially a mathematical representation based on data such as price (high, low, open, close, etc.) or the volume of a security to predict price trends [2]. These indicators are tools used in trading charts to help make market analysis clearer for traders. Technical indicators serve different purposes and are divided into four categories: trend, momentum, volume, and volatility. For instance, momentum indicators measure the speed and magnitude of price movement, volume indicators relate to price and volume, and volatility indicators show the degree of price fluctuation. While volume and volatility indicators do not directly show trends, they can indicate stock trends when combined with movement indicators. Various studies have been conducted on technical indicators, but no comprehensive analysis has been performed to determine which indicators work best with specific machine learning (ML) methods. Therefore, it is crucial to understand which combinations of indicators and machine learning techniques can enhance stock market predictions [3]. Considering the long history of stock prediction, most studies have tried to use technical indicators for stock forecasting [2]. Many technical indicators have been developed, and new types are still being created by traders to achieve better results. However, stock market predictions cannot be made with a single indicator alone, and an appropriate combination of indicators should be used for prediction. The number of technical indicators used to predict stock prices is growing. Some of these indicators can create conflicting signals, so it is essential to choose the right combination of indicators when making stock investment decisions. Due to the vast number of possible combinations, it is impossible to use all of them. Therefore, having algorithms that can generate suitable combinations of technical indicators is crucial.

Feature selection is an essential step in improving model performance. Feature selection is the process of selecting the most relevant and important features from a large set of features, improving computational efficiency, and reducing model generalization errors [13]. This process also helps to prevent overfitting and mitigate the curse of dimensionality [4]. Various methods are available for selecting the best features [5]. These methods are mainly divided into two types: supervised Feature Selection and unsupervised feature selection. Supervised feature selection methods consider the target variable and are used for labeled datasets, while unsupervised methods ignore the target variable and are suitable for unlabeled datasets [6]. Supervised feature selection methods include three main approaches: filter methods [20], wrapper methods, and embedded methods. The wrapper method was used by Kohavi and Jan in 1997 to select a subset of the set of all features [8]. For each of the subsets, a supervised learning model (e.g., classification) is fitted. Then these subsets are evaluated with a performance measure calculated based on the resulting model (such as classification accuracy). In these methods are used, simple algorithms such as Kitler's article [9], That is, Greedy Sequential Searches, more complex algorithms such as the Recursive Feature Elimination in the article by Huang et al. [10]. As well as Evolutionary and Swarm Intelligence Algorithms in the article by Xue et al. [11]

and Brezočnik et al. [12] has been used for feature selection. The flowchart of wrapper methods is shown in Figure 1.

Wrapper methods, in general, consist of three main components: (1) a search algorithm, (2) a fitness function, and (3) an inductive algorithm [36]. The wrapper routine terminates when a predefined number of iterations is reached or the desired number of features is selected (i.e., greedy search). In wrapper methods, the feature selection process is based on a specific machine learning algorithm that is tried to be placed on a given data set. It follows a greedy search approach by evaluating all possible combinations of features against an evaluation criterion. The evaluation criteria are simply the performance criteria that depends on the type of problem. For example, the regression evaluation criteria can be MSE, R-squared, MAE. Similarly, for classification, the evaluation criterion can be accuracy, precision, recall, F1-score, etc. Finally, it selects a combination of features that achieve optimal results for the specified machine learning algorithm [37].

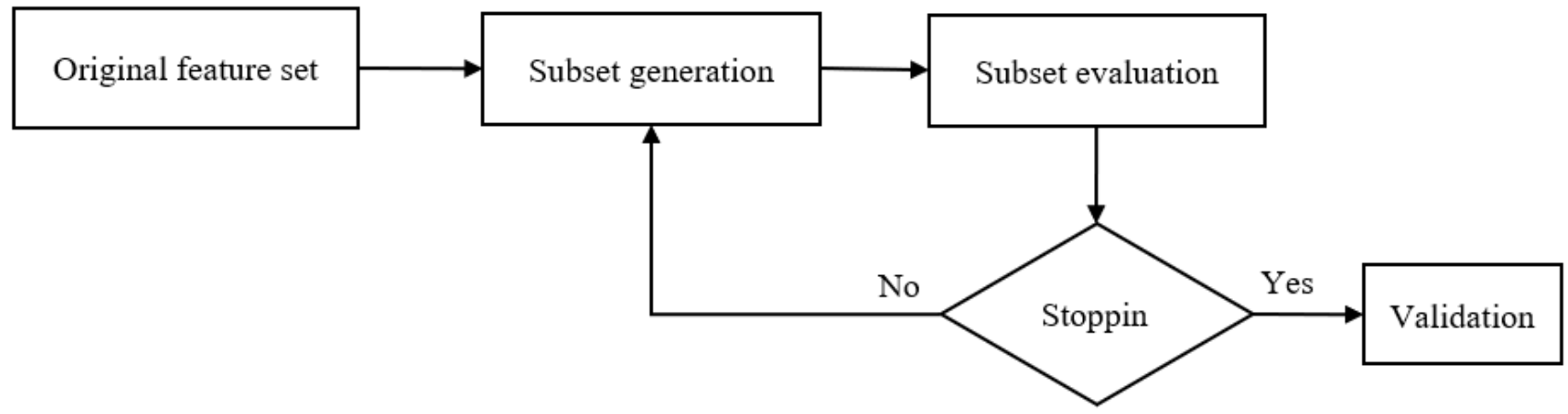

**Fig. 1** The flowchart of wrapper methods [8]

The wrapper methods include a learning algorithm as a black box and using its predictive performance to evaluate the usefulness of a subset of variables. The most common techniques used under wrapper methods are:

**Forward Selection:** Forward Selection starts with an empty set of features and is an iterative process. In each iteration, a feature is added and it is checked whether this feature improves the performance of model or not. Iteration continues until adding a new variable or feature improves the performance of the model [48].

**Backward Selection:** Backward Selection is an iterative approach like Forward Selection, but it is the opposite of Forward Selection. Backward Selection starts by considering all the features and removing the features that have the least performance with the method and continues until removing the features does not improve the performance of the model [4].

In this study, we use machine learning techniques, specifically regression models combined with feature selection methods, to improve prediction accuracy. For this purpose, 10 different regression methods will be examined. By evaluating all types of technical indicators with various regression methods, we aim to predict stock market prices. This research aims to assist traders in maximizing profits or minimizing potential losses through the identification of optimal technical indicators. Thus, the primary objective is to select the model that performs best in predicting stock prices using various indicators. The secondary goal is to determine the most effective indicators for different machine learning methods to enhance stock market predictions. This study specifically focuses on underexplored indicators that have received less attention from researchers. To date, there has been limited exploration into the optimal combination of technical indicators for stock market analysis.

The novelties of this research are as follows:

(a) A novel method for using time windows with regression models, introduced for the first time in this study. By creating a time window, many features are generated, transforming a two-dimensional dataset into a three-dimensional one. Since regression methods typically require two-dimensional input, our approach converts it back into a processable two-dimensional format.

(b) Improvement in both prediction accuracy and speed of regression methods through the selection of optimal features. This is achieved for the first time by using a specific combination of features, which has a significant impact on the model's performance.

(c) The utilization of numerous novel technical indicators, contributing to higher accuracy in stock price predictions. These technical indicators are effectively used in combination with regression methods for the first time in this research.

(d) The use of the most effective combination of indicators and regression methods to predict stock market prices, resulting in improved performance. This particular combination is implemented for the first time in this study.

(e) The results show that the proper combination of the proposed indicators and regression methods leads to high accuracy in predicting stock closing prices. This unique combination and approach are used for the first time in this research.

the innovations in your work clearer and more specific.

Current research fills gaps in previous studies by:

1. Using a broader set of technical indicators (123 indicators) and 10 regression methods, compared to the limited approaches used in prior work.
2. Analyzing the impact of time window size on prediction accuracy, with a 3-day window being identified as optimal.
3. Evaluating models using 5 error metrics, providing a more comprehensive measure of performance than previous studies.
4. Improving prediction accuracy with models like MLPSF, SVRSF, LRSF, and RISF, showing significant improvements over traditional models.
5. Identifying effective technical indicators (e.g., Squeeze_pro, Percentage Price Oscillator) that weren't explored in-depth before.

Overall, current work offers a more precise and comprehensive approach to stock price prediction by combining optimal feature selection and regression methods.

## 2 Literature Review

One of the effective ways that traders use to predict the stock market is to take the help of technical analysis. The basis of this analysis focuses on the price and history of various financial assets. This analysis with the help of charts can provide a good prediction of the future of a financial asset. In technical analysis, current prices and expectations are checked with the past of that price. What is important in technical analysis to predict the stock market is time. In technical analysis, past data of the market is used to predict the direction of prices and it depends on statistical methods to identify patterns [21, 22].

To solve the problem of stock closing price prediction, Xu et al. [23] proposed two new clustering-based prediction models based on two-stage prediction models. They used technical indicators (SMA, EMA, MOM, MACD, RSI, etc.) and machine learning models like SVM and random forest. They employed K-Means clustering [24, 25] on technical indicators to divide stocks into clusters, then used ensemble learning to improve prediction accuracy. The model was tested on four Chinese stocks, with evaluation using MAPE, MAE, RMSE, and MSE. The results demonstrated improved prediction accuracy through the combined K-Means and ensemble learning model.

Fazeli and Houghten [26] used deep learning, specifically LSTM networks [27, 28], to predict stock trends, utilizing technical indicators such as RSI, Williams %R, and volatility. The data were selected from one of the S&P 500 companies, and a backtesting approach was applied to measure real-world performance. The model showed significant improvement in stock return predictions and outperformed traditional methods like buying and holding assets, especially for Apple and Intel. Hyperparameter optimization was performed using Talos. The stock return increased from a negative value to 6.67%. A positive ROI was obtained for all four stocks. Compared to buying and holding assets, the performance of the proposed approach was significantly better for Apple and Intel, but worse for Microsoft and Google. Over the same period, the S&P 500's overall performance was +0.05%.

Xiao et al. [29] proposed a stock price prediction model based on ARI-MA-LS-SVM, which combines ensemble auto-regression moving average (ARI-MA) and least squares support vector machine (LS-SVM). The model uses various technical indicators such as moving averages, KDJ, and trading volumes to predict the next day's closing price. After feature reduction and training on selected sample groups, the model showed high accuracy and low error, outperforming LS-SVM and RS-SVM models.

Zhen et al. [30] proposed a combined Particle Swarm Optimization-Support Vector Machine (PSOSVM) model to overcome the limitations of Support Vector Machines (SVMs) when dealing with noisy and high-dimensional data. PSO is used for feature selection and parameter optimization, while SVM predicts stock trends. The model was tested on Malaysian stock data over 17 years, using 15 technical indicators. The PSOSVM approach demonstrated better prediction accuracy by removing irrelevant features and optimizing SVM parameters.

Mohanty et al. [31] introduced a modified multi-layer extreme learning machine (ELM) combined with a random auto encoder (AE) for stock market prediction. The AE-KELM model was tested using financial data (OHLC) for various markets, including banks and stock exchanges in China, India, and the US. The model showed promising results in comparison with other techniques, using evaluation methods like MAPE, MAE, and RMSE. The study found that the polynomial kernel function provided the best prediction accuracy. Accuracy improved by approximately 5% to 15%. Performance depends on the appropriate combination of models.

Hoseinzade and Haratizadeh [32] proposed CNNpred, a convolutional neural network (CNN) model designed to extract features from multiple financial variables and predict stock market trends. They used data from major US stock indices (S&P 500, NASDAQ, etc.), exchange rates, commodity prices, and other financial indicators. CNNpred has two versions: 2D-CNNpred, which uses a general model for multiple markets, and 3D-CNNpred, which trains specific models for individual markets. The model applies feature extraction and CNN layers to predict future market fluctuations.

Mishra et al. [33] used multiple linear regression with backward elimination to forecast the TCS stock index. The model utilized independent variables such as opening price, highest price, and lowest price to predict the closing price. The research showed that all independent variables had strong correlations with the dependent variable, and the model achieved an R-square value of 1, indicating excellent prediction accuracy.

Patil proposed [34] the Deep Recurrent Rider LSTM model, a combination of Rider Deep LSTM [35] and Deep RNN [36, 37] classifiers, trained using the SCSO, which is a combination of SSO [38] and CSA [39], algorithm. The model was applied to stock markets of two companies, Reliance Communications and Relaxo Footwear, using data from 2019 to 2021. Feature extraction [40], data augmentation using bootstrap, and error minimization techniques were used to improve prediction accuracy. The model produced predictions based on the least error between the classifiers.

Naik and Mohan [46] used 33 combinations of technical indicators to predict short-term stock movements for ICICI Bank and State Bank of India (2008-2018). They employed the Borota feature selection technique to identify the most relevant features for stock price prediction. The Artificial Neural Network (ANN) with three layers and a sigmoid activation function was used. The ANN model was trained with gradient descent to minimize global error, achieving a performance improvement of 12% over existing methods. Evaluation was done using Mean Absolute Error (MAE) and Root Mean Square Error (RMSE).

Ji et al. [48] proposed a method for improving the accuracy of technical indicators by denoising stock price data using wavelet basis functions. The study focused on four stock markets (SSEC, Hong Kong, S&P 500, and DJI) and reviewed 18 technical indicators. They employed variable time window sizes (3, 5, 10, 15, 30, 45, and 60 days) and aimed to predict stock prices three days ahead. The results showed that denoising the data improved the accuracy of the model.

Haq et al. [49] focused on classifying stock price movement using 44 technical indicators from daily data of 88 stocks. They trained logistic regression, support vector machines, and random forests to predict stock movement. The accuracy and the Mathius Correlation Coefficient (MCC) were used as evaluation metrics, and the results indicated that their model performed well in predicting stock movements.

Peng et al. [51] analyzed 124 technical indicators and used feature selection techniques (SFS, SBS, Lasso) to predict stock prices. They employed deep neural networks with various architectures (3, 5, and 7 hidden layers) trained on data from seven global markets (2008-2019). The networks were optimized with the Adam algorithm over 400 epochs, achieving prediction accuracies between 50% and 65% for all seven markets and 48 compositions. The feature selection process significantly impacted the model's performance.

This study explores the application of regression methods with feature selection and time windowing in predicting system performance. The materials and methods used in previous works are discussed below in Table 1.

**Table 1** Comparison of Previous Works

| Reference | Support Vector Regression | Decision Tree Regression | Random Forest Regression | Neural Network and Multi-Layered Perceptron Regression | Linear Regression | K-Nearest Neighbor Regression | Adaboost Regression | Gradient Boosting Regression | Lasso Regression | Ridge Regression | Feature Selection | Technical Indicator | MAPE | RMSE | MAE | MAE | $R^2$ | Other Metrics |
|---|---|---|---|---|---|---|---|---|---|---|---|---|---|---|---|---|---|---|
| Dogan et al. [7] | * | * | * | * | | * | * | | | | | | | | | | | Accuracy |
| Wang et al. [14] | * | | * | * | | | | * | | | | | * | | | | | |
| Patel et al. [15] | * | * | | | | | | | | | | | | | | | | Accuracy |
| Xu et al. [23] | * | | * | | | | | | | | | * | * | * | * | * | * | |
| Zhen et al. [30] | * | | | | | | | | | | * | * | | | | | | Accuracy |
| Mohanty et al. [31] | | | | * | | | | | | | * | * | * | * | * | | | |
| Hoseinzade & Haratizadeh [32] | | | | * | | | | | | | | * | * | * | * | | | F1-Score Accuracy |
| Guo et al. [16] | | | | * | | | | | | | | | * | * | | | | R |
| Mishra et al. [33] | | | | * | * | | | | | | * | | | | | | * | Accuracy |
| Naik et al. [46] | | | | * | | | | | | | | | | * | * | | | |
| Ji et al. [48] | | | * | | | | | | | | | * | | | | | | F1-Score |
| Haq et al. [49] | * | | * | | * | | | | | | | * | | | | | | MCC |
| Peng et al. [51] | | | | * | | | | | * | | * | | * | | | | | Precision, Recall Accuracy F1-Score |
| Current Research | * | * | * | * | * | * | * | * | * | * | * | * | * | * | * | * | * | |

## 3 Proposed Method

The process begins with selecting the most suitable combination of indicators for predicting stock market prices. Subsequently, the proposed models, combined with these optimized indicator combinations, will be employed to forecast stock market prices. To ensure a precise evaluation of the models, separate datasets are utilized one for selecting the best indicators and another for predicting stock market prices [37].

The primary objective of this study is to identify the most effective model for predicting stock market prices with different indicators, achieving the least prediction error. The second objective is to determine the optimal combination of indicators using various machine learning methods to enhance stock market price prediction accuracy. The third objective involves utilizing a time-window approach for regression-based predictions. By implementing the time window, a two-dimensional dataset is transformed into a three-dimensional dataset, which is then converted back into a two-dimensional input compatible with regression methods using the proposed technique. The findings of this study demonstrate how selecting the right combinations of indicators can significantly improve stock price prediction. Predictions are evaluated using error-based performance criteria, allowing for the identification of the best methods for stock price forecasting [52].

The dataset used in this research consists of Apple's stock data from January 1, 2010, to December 31, 2022, sourced from Yahoo Finance. The raw data contains six columns: (1) Open, (2) High, (3) Low, (4) Close, (5) Volume, and (6) Adjusted Close. Among these, the closing price is the dependent variable, while the remaining columns are used exclusively to calculate technical indicators and are later excluded from the dataset. Therefore, the independent variables in this study are the derived technical indicators.

This research applies wrapper feature selection techniques in combination with various regression methods to identify the optimal combination of technical indicators. Wrapper methods are used to select the best indicators based on different models, thereby enhancing the accuracy of stock market price predictions. A comprehensive evaluation of all technical indicators using prominent machine learning methods has been conducted, representing a novel approach that has not been previously addressed.

In total, 123 technical indicators have been examined, as detailed in Table 2. All indicators utilized in this study are calculated using the Pandas TA library.

This research is conducted in two stages, in the first stage, Forward Selection and Backward Selection methods are applied to various regression models. These models are then trained, and the best combination of technical indicators is identified based on evaluation criteria. In the second stage, a new dataset is created using the subset of indicators determined in the first stage. After training the models with the training set, they are employed to predict the testing set. The overall framework for selecting the optimal indicators for stock market price prediction using machine learning methods is illustrated in Figure 2.

**Table 2** The definition of the 123 indicators in the data sets

| No. | Name | No. | Name | No. | Name | No. | Name | No. | Name |
|---|---|---|---|---|---|---|---|---|---|
| 1 | aberration | 26 | decreasing | 51 | kc | 76 | psl | 101 | supertrend |
| 2 | apo | 27 | dema | 52 | kvo | 77 | qstick | 102 | Sine wma |
| 3 | accbands | 28 | dpo | 53 | kst | 78 | qqe | 103 | swma |
| 4 | ad | 29 | dm | 54 | linreg | 79 | roc | 104 | t3 |
| 5 | adosc | 30 | donchian | 55 | massi | 80 | rsi | 105 | td_seq |

| 6 | alma | 31 | eom | 56 | mcgd | 81 | rsx | 106 | tema |
|---|---|---|---|---|---|---|---|---|---|
| 7 | aobv | 32 | ebsw | 57 | midpoint | 82 | rvgi | 107 | thermo |
| 8 | aroon | 33 | er | 58 | midprice | 83 | rvi | 108 | trima |
| 9 | adx | 34 | ssf | 59 | mom | 84 | rma | 109 | trix |
| 10 | atr | 35 | eri | 60 | mfi | 85 | zscore | 110 | true_range |
| 11 | ao | 36 | efi | 61 | macd | 86 | stdev | 111 | tsi |
| 12 | bop | 37 | ema | 62 | nvi | 87 | kurtosis | 112 | ttm_trend |
| 13 | bias | 38 | fwma | 63 | natr | 88 | mad | 113 | ui |
| 14 | bbands | 39 | fisher | 64 | ohlc4 | 89 | median | 114 | uo |
| 15 | brar | 40 | hilo | 65 | obv | 90 | quantile | 115 | vhf |
| 16 | cg | 41 | hl2 | 66 | psar | 91 | skew | 116 | vidya |
| 17 | cmf | 42 | hlc3 | 67 | pwma | 92 | variance | 117 | vp |
| 18 | cfo | 43 | hwc | 68 | ppo | 93 | stc | 118 | vwap |
| 19 | cksp | 44 | hma | 69 | pvo | 94 | sma | 119 | vortex |
| 20 | cmo | 45 | hwma | 70 | pvi | 95 | slope | 120 | wcp |
| 21 | chop | 46 | ichimoku | 71 | pgo | 96 | smi | 121 | willr |
| 22 | cci | 47 | increasing | 72 | pdist | 97 | squeeze | 122 | wma |
| 23 | coppock | 48 | amat | 73 | pvr | 98 | squeeze_pro | 123 | zlma |
| 24 | cti | 49 | kama | 74 | pvt | 99 | stoch | | |
| 25 | decay | 50 | kdj | 75 | pvol | 100 | stochrsi | | |

Figure 2 illustrates the process. First, the stock data of Apple Inc. (AAPL) is collected from Yahoo Finance. Next, various technical indicators (listed in Table 1) are computed, and a new dataset is created that includes these technical indicators alongside the stock data. The collected data then undergoes preprocessing to address missing values, which are replaced with the mean of the available data. Each feature in the dataset is normalized using a Min-Max Scaler to bring the data to a uniform scale. Additionally, the data is analyzed to evaluate the relevance of features, allowing for the selection of important features that contribute to accurate stock price prediction. The feature selection stage focuses on identifying the best technical indicators for use with regression methods. A detailed flowchart of the stock market price prediction process is presented in Figure 3.

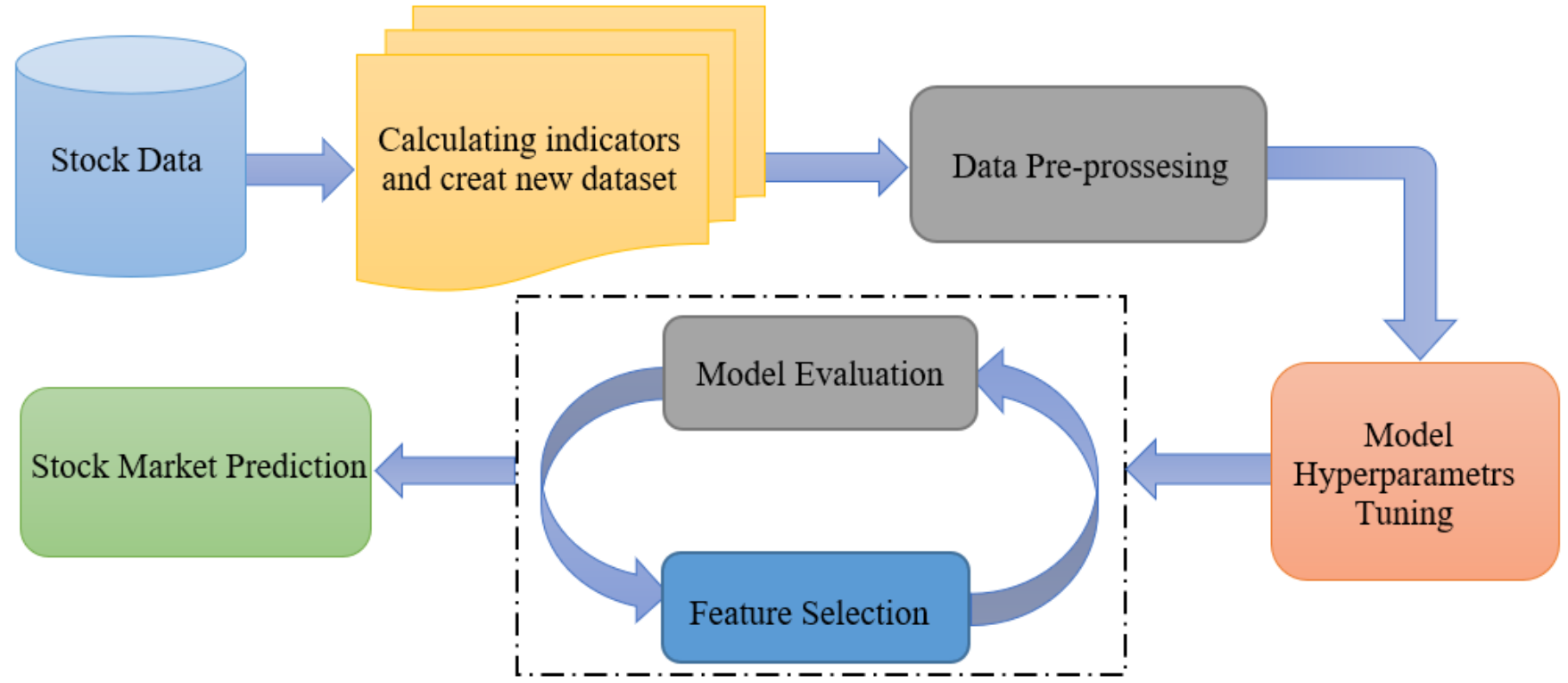

**Fig. 2** General framework for choosing the best combination of indicators for stock market prediction

The primary objective of this study is to identify the model that delivers the best performance for prediction using various indicators. The secondary goal is to determine the most effective indicators for different machine learning methods to enhance stock market price prediction accuracy.

As in the study conducted by Peng et al. [51], the dataset in this research is divided into two parts. The first part contains the historical data of Apple Inc. (AAPL) stock from 01/01/2010 to 31/12/2013, consisting of 1,006 samples. This portion is used for feature selection. The second part includes historical data from 01/01/2014 to 31/12/2022, comprising 2,266 samples, and is used for stock market price predictions. For both parts of the dataset, a 70-30 ratio is applied to split the data into training and testing sets. The training set is utilized to train the model, while the testing set serves as unseen data where the algorithm's predicted outputs are compared with the actual outputs to evaluate performance.

However, dividing the dataset into a training and testing set often introduces high variance in the results, particularly when the dataset size is limited. This can lead to varying outcomes when the experiment is repeated. To mitigate this issue, K-fold cross-validation is employed. Cross-validation builds and evaluates multiple models on different subsets of the dataset, generating several performance metrics. These metrics are then averaged to provide a more reliable estimate of the method’s overall performance.

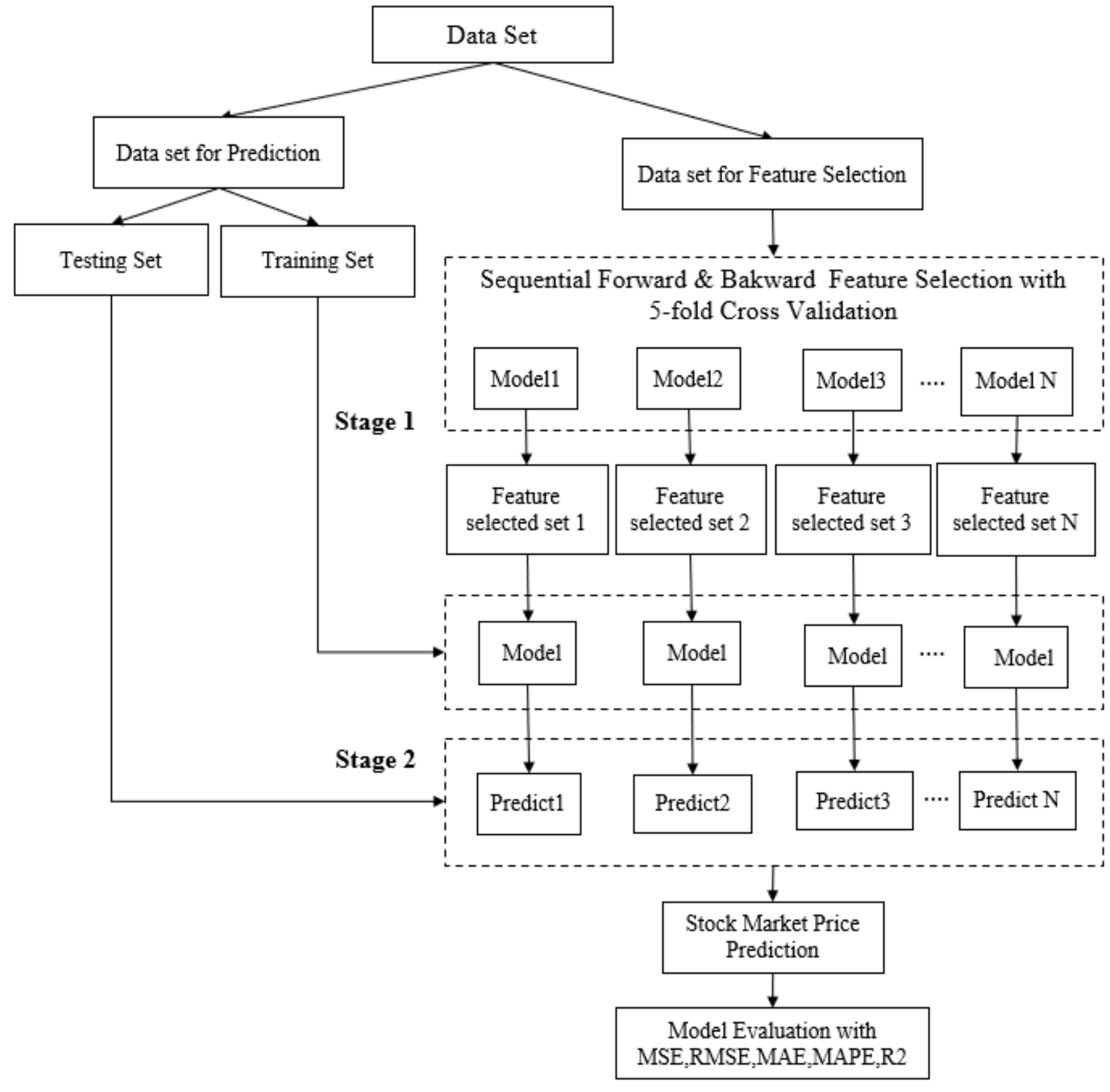

**Fig. 3** Flowchart of the proposed stock prediction model

## 3.1 Using time window transformation via flattening

The time window is a way to convert a time series prediction problem into a supervised learning problem. Here, like Ji et al. [48], a 3-day time window is intended for the data set and a new data set is created (Figure 5, Equation (1)). The goal here is to predict the price after 3 days. The 3-day time window will increase the features in the data set. Therefore, after creating a time window, the new data set should be applied to a variety of machine learning methods as input. The time window transforms the two-dimensional dataset into a three-dimensional one. However, since regression methods require a two-dimensional input, we apply a flattening technique to convert the data back into a processable two-dimensional format. In accordance with Equation (2) and as shown in Figure 6, this new data set will be converted into inputs usable by machine learning methods.

Here $S_i$ is the time window $i$, $w$ is the time window size, $day_j$ is the day $j$ in the data set, $k$ is the number of features (indicators) [52].

$$s_i = \{day_j, day_{j+1}, \dots day_{w+j-1}\}, \qquad i = j-1, \qquad j = 1, \dots, n-w+1 \tag{1}$$

$F_{k,j}^{(i)}$ is the value of the indicator $k$ for the day $j$ in window $i$.

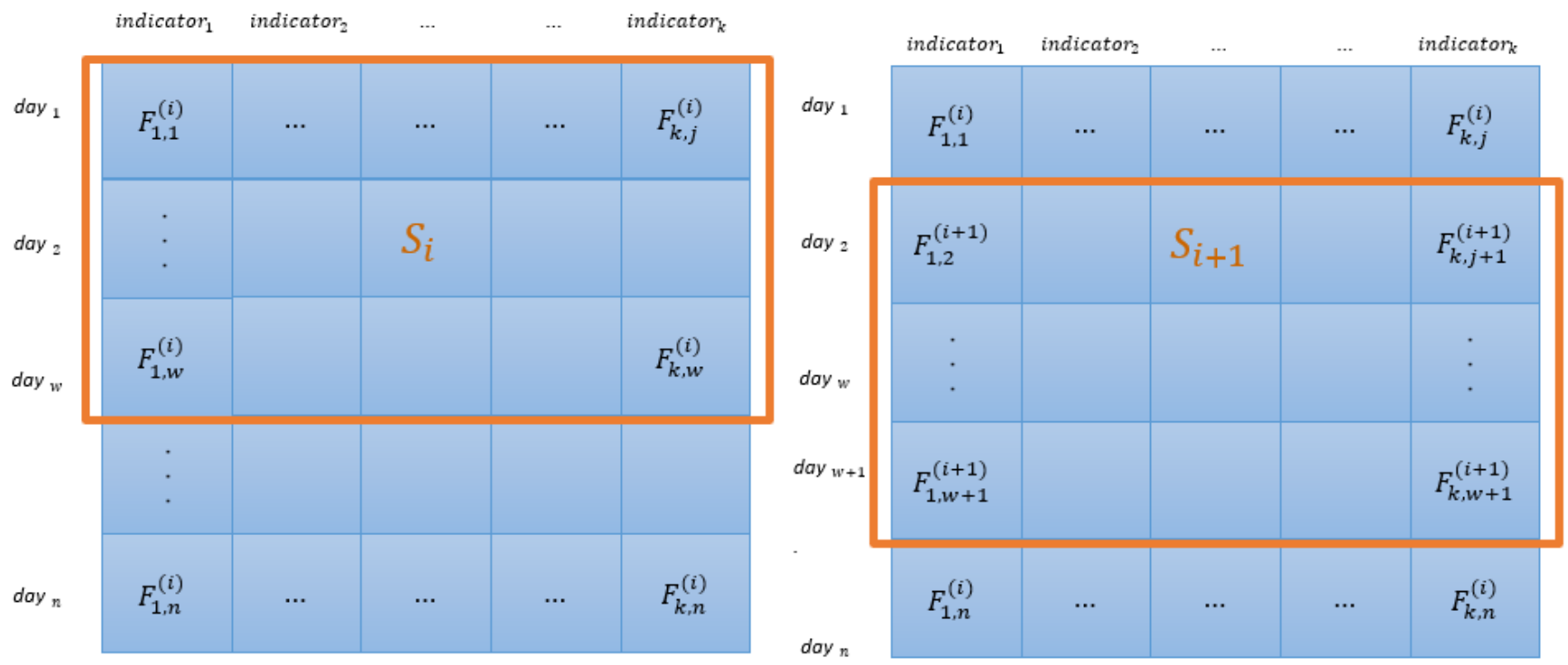

**Fig. 5** Create a time window for data set

Data for each time window is converted to a data point called $C_{i,j}$. $C_{i,j}$ contains the value of indicators of the day $j$ in the time window $s_i$ [52].

$$C_{i,j} = \{F_{1,1}^{(i)}, F_{2,1}^{(i)}, \dots, F_{k,j}^{(i)}\} \tag{2}$$

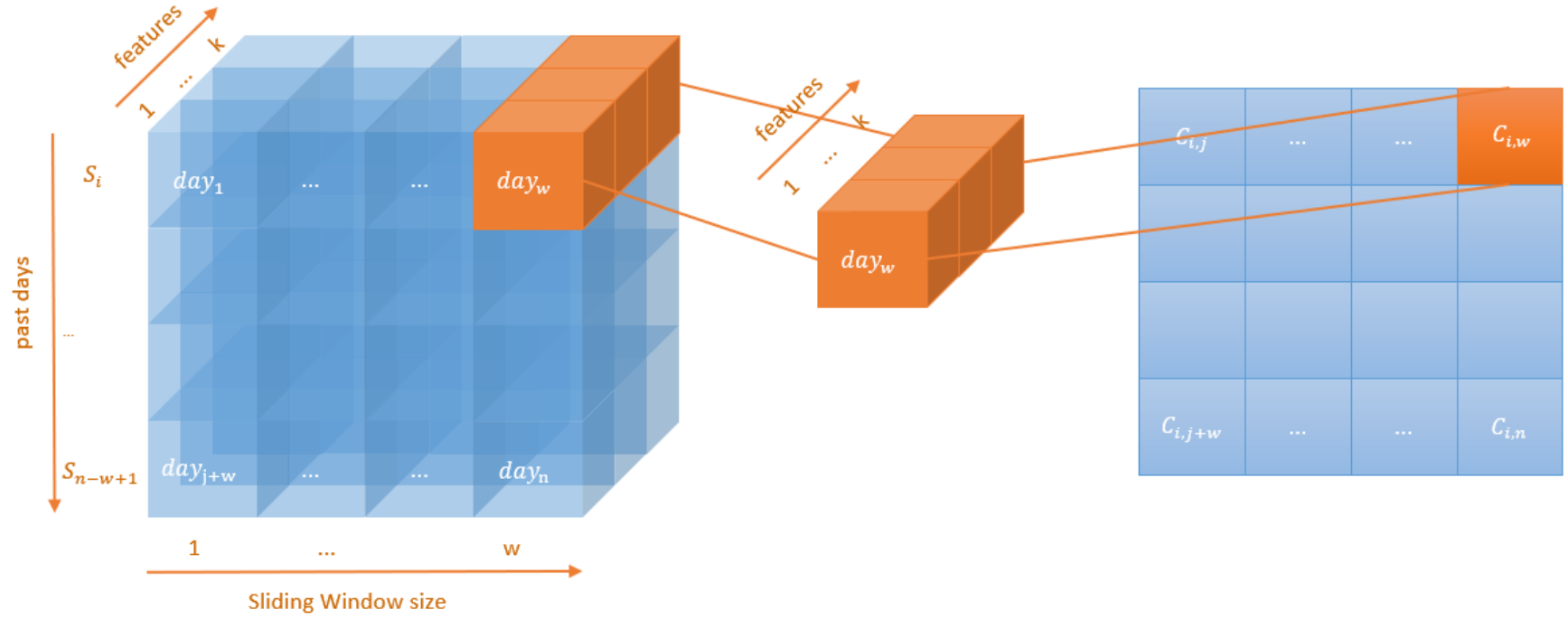

**Fig. 6** Using time window transformation via flattening

In this research, six groups of estimators from the scikit-learn library have been examined, which include a wide range of estimators, estimators such as; Simple linear estimator (linear regression, Ridge regression and Lasso regression), based on decision tree (decision tree regression), based on nearest neighbor (K-nearest neighbor regression), based on neural network (multilayer perceptron regression) , based on support vector machine (support vector regression) and group based (i.e. AdaBoost regression, Gradient Boosting regression and random forest).

(1) Linear Regression (LR): LR is a type of linear predictive function in which the dependent variable is predicted as a linear combination of independent variables. Each of the independent variables is multiplied by the coefficient obtained for that variable in the estimation process; The final answer will be the sum of the products plus a constant value, which is also obtained in the estimation process [47].

(2) Ridge Regression: The loss function in the Ridge regression is the function of the Linear least squares and is regulated by l2-norm [19].

(3) Lasso Regression: The Lasso regression is a good way to model the response variable based on the lowest and most appropriate number of independent variables and tries to separate the more appropriate variables from the rest of the variables and provide a simpler model [41].

(4) Decision Tree Regression (DTR): The decision tree is a tree function consisting of several decision-making nodes. Every non-leaf node is a feature. In the decision tree, the data feature is shown in the inner nodes of the branches and their result in the leaf of each branch [44].

(5) K-Nearest Neighbor Regression (KNN): Neighbor-based regression can be used in cases where variables are continuous. The K-Nearest neighbor regression implements learning based on (k) of the nearest neighbors of each point and uses uniform weights. In some cases, the weight of the points can be beneficial, so that the nearest points are more likely to be involved in the regression than the distant points [42].

(6) Multi-Layered Perceptron Regression (MLP): MLP consists of an input layer, hidden layers, and an output layer. This model is trained with a back propagation algorithm. MLP optimizes the square error using LBFGS or Stochastic Gradient Descent. Activation functions for hidden layers include; Logistic, Tanh, Softsign, Softplus, Sigmoid, Relu, Exponential, Selu, Elu, Identity, LeakyReLU. The MLP is repeatedly trained because at each time the partial derivatives of the loss function are calculated according to the model parameters to update the parameters. It can also add a regularization term to the loss function, which reduces the model parameters to prevent overfitting. The square error uses the loss function and the output of the set is from continuous values [42].

(7) Support Vector Regression (SVR): The support vector machine (SVM) is a generalized linear classification (supervised learning) that categorizes the data into binary categories. The boundary of that cloud decision is the maximum margin of the sample [38]. In the year 1996, Vapnic and his colleagues proposed a copy of SVM that performs regression instead of classification. This item is known as SVR. The support vector regression is highly accurate for predicting the stock market, but because of the time consuming, its parameters are not used. Failure to accurately adjust its parameters can lead to a time-consuming method that diverts researchers' attention to more use of this technique [43].

(8) Adaboost Regression (ADA): The Adaboost regression is an ensemble machine learning method that starts with the fit of a regression on the original data set and then places other versions of the regression on the same data set, but the weight of the samples is adjusted according to the current prediction error. As a result, the next level regression focuses more on the defects of previous regression [49].

(9) Gradient Boosting Regression (GBR): Gradient boosting is an ensemble machine learning method used for regression and classification [18]. The Gradient boosting is a linear combination of weak models created to create a strong final model [45].

(10) Random Forest Regression (RFR): The random forest creates an ensemble model with basic decision -making trees [50]. The random forest makes several decisions and integrates them to make more accurate and sustainable predictions. One of the benefits of random forest is its usability for both categorization and regression issues [17].

Table 3 summarizes the abbreviations used for each regression method combined with SFS and SBS feature selection techniques.

**Table 3** Abbreviations of Regression models combined with Feature Selection methods

| Abbreviation | Full Method Description |
|---|---|
| **LRSF** | Linear Regression + Sequential Forward Selection (SFS) |
| **LRSB** | Linear Regression + Sequential Backward Selection (SBS) |
| **RISF** | Ridge Regression + SFS |
| **RISB** | Ridge Regression + SBS |
| **LOSF** | Lasso Regression + SFS |
| **LOSB** | Lasso Regression + SBS |
| **DTRSF** | Decision Tree Regression + SFS |
| **DTRSB** | Decision Tree Regression + SBS |
| **KNSF** | K-Nearest Neighbor Regression + SFS |
| **KNSB** | K-Nearest Neighbor Regression + SBS |
| **MLPSF** | Multilayer Perceptron Regression + SFS |
| **MLPSB** | Multilayer Perceptron Regression + SBS |
| **SVRSF** | Support Vector Regression + SFS |
| **SVRSB** | Support Vector Regression + SBS |
| **ADASF** | AdaBoost Regression + SFS |
| **ADASB** | AdaBoost Regression + SBS |
| **GBSF** | Gradient Boosting Regression + SFS |
| **GBSB** | Gradient Boosting Regression + SBS |
| **RFRSF** | Random Forest Regression + SFS |
| **RFRSB** | Random Forest Regression + SBS |

In this study, grid search was employed to identify the optimal hyperparameters for the models. It is important to note that not all hyperparameters hold the same level of significance for model performance. The evaluation of each hyperparameter set was conducted using K-fold cross-validation, which divides the training dataset into K subsets. Specifically, hyperparameter tuning in this study was performed with 10-fold cross-validation, repeated 3 times to ensure reliable results.

The best hyperparameters identified and used in this research are summarized in Table 4.

**Table 4** Hyperparameters tuning

| Model | Best Hyperparameters |
|---|---|
| **LR** | copy_X= True, fit_intercept= True, n_jobs= None, positive= False |
| **LASSO** | Alpha=0.1, max_iter= 200 |
| **Ridge** | Alpha= 0, fit_intercept= True, solver='svd' |

| DTR | Criterion= 'squared_error', max_depth= 9, min_samples_leaf= 2 |
|---|---|
| **KNN** | leaf_size= 10, metric= 'manhattan', n_neighbors=2, weights='distance' |
| **MLP** | Activation='logistic', alpha= 1, hidden_layer_sizes= 50, learning_rate= 'constant', learning_rate_init= 0.0001, max_iter= 2000, random_state= 1, solver= 'lbfgs' |
| **SVR** | C= 0.1, gamma= 0.1, kernel= 'rbf' |
| **ADA** | learning_rate= 0.1, loss= 'square', n_estimators= 2000, random_state=1 |
| **GBR** | alpha= 0.1, criterion='squared_error', learning_rate= 0.1, loss= 'squared_error', max_leaf_nodes= 30, n_estimators=2000 |
| **RFR** | max_features= 20, n_estimators= 1000, Criterion='squared_error', min_samples_split=2 |

### 3.2 Evaluation Criteria

In regression problems, one or more target characteristics are predicted based on the input characteristics of the model. Unlike classification problems, the target feature is continuous. To evaluate the accuracy of the model, we use $R^2$, MSE, RMSE, MAE and MAPE, which are common metrics in regression tasks.

Coefficient of determination or score $R^2$ measures how well the predicted values fit the actual data. It ranges from 0 to 1. A higher $R^2$ means the model explains more variance in the target variable and is calculate by Equation (3) [53].

$$R^2 = 1 - \frac{\sum_{i=1}^{N} (y_i - \hat{y}_i)}{\sum_{i=1}^{N} (y_i - \bar{y}_i)} \tag{3}$$

Here $y_i$ is the actual value, $\hat{y}_i$ is the predicted value, $N$ is the total number of data point.

Mean Square Error (MSE) is the average of the squared differences between actual and predicted values. It heavily penalizes large errors and shown in Equation (4) [53].

$$MSE = \frac{1}{N} \sum_{i=1}^{N} (y_i - \hat{y}_i)^2 \tag{4}$$

Root Mean Squared Error (RMSE) is the square root of MSE. It's in the same units as the target variable and is easier to interpret than MSE and shown in Equation (5) [53].

$$RMSE = \sqrt{MSE(y, \hat{y})} \tag{5}$$

Mean Absolute Error (MAE) is the average of the absolute differences between actual and predicted values. It treats all errors equally and is presented in Equation (6) [53].

$$MAE = \frac{1}{N} \sum_{i=1}^{N} |y_i - \hat{y}_i| \tag{6}$$

Mean Absolute Percentage Error (MAPE) is the average of the absolute percentage errors between actual and predicted values. It is expressed as a percentage and shown in Equation (7) [53].

$$MAPE = \frac{100}{N} \sum_{i=1}^{N} \left| \frac{y_i - \hat{y}_i}{y_i} \right| \tag{7}$$

# 4 Results

In this study Sequential Forward Selection (SFS) and Sequential Backward Selection (SBS) used with Linear regression, Ridge regression, Lasso regression, Decision Tree regression, K-Nearest Neighbor regression, Multilayer Perceptron regression, Support Vector Regression, Adaboost Regression, Gradient Boosting Regression and Random Forest Regression. A total of 10 machine learning models are evaluated, each with two methods of selection and each method with 5 evaluation criteria stated and finally, 100 algorithms are implemented. Sequential Forward Selection methods (SFS) and Sequential Backward Selection (SBS) with any method and any evaluation criteria, each is executed with cross validation equal to 5 (CV=5) and select the feature set with the highest rating as the best set. The features selected by each method are in Table 4 in appendix. In the experiments, we used the technical indicators selected as the input features per proposed method with 3-day time windows. Each method selects different indicators by different evaluation criteria by SFS and SBS methods. Also, each method when selecting the feature is correlated with different indicators, and these indicators will improve the stock market prediction.

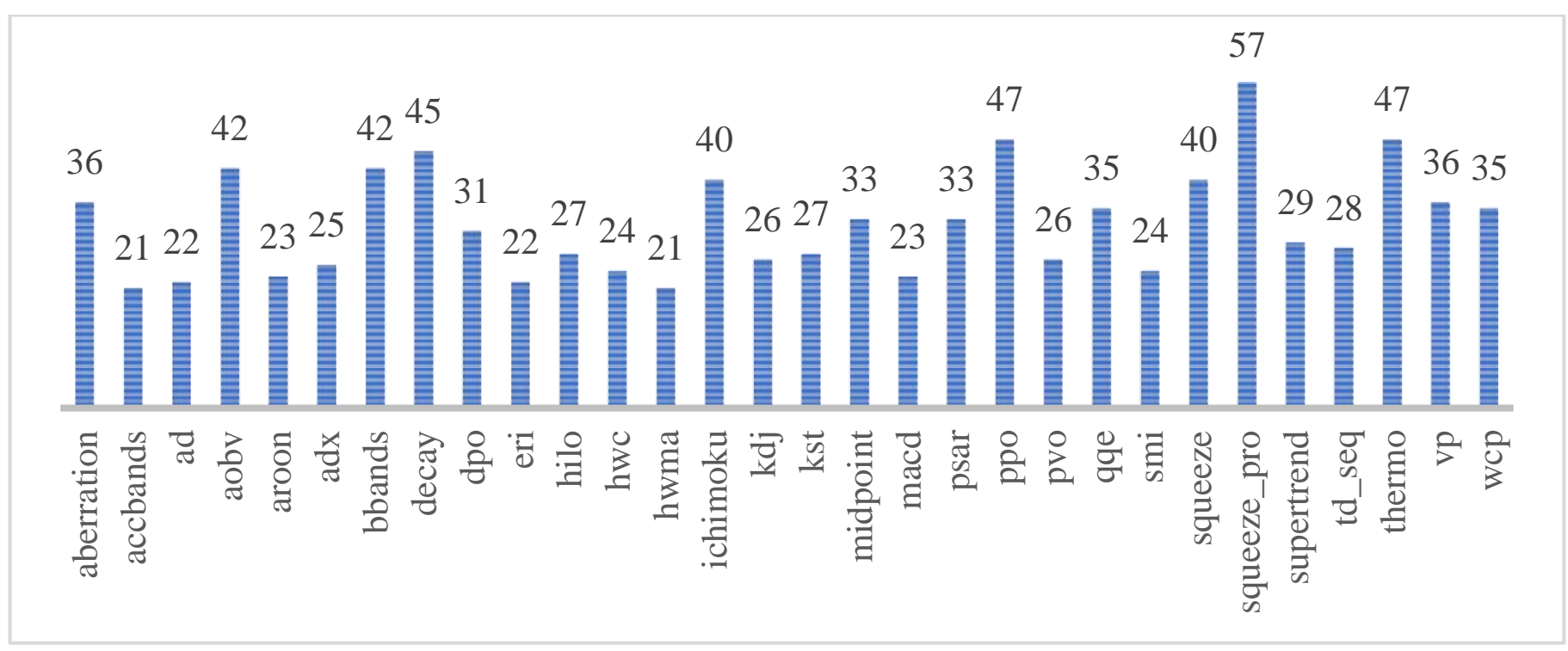

**Fig. 7** Percentage of use of 30 top selected indicators with 10 regression methods

Based on Figure 7, the Squeeze_pro indicator with 57% is used to better predict the market. Also, the most commonly used indicators to predict the stock market are; the Percentage Price Oscillator indicator with 47%, The Thermo indicator with 47%, Decay indicator with 45%, Archer On-Balance Volume with 42%, Bollinger Bands indicator with 42%, Squeeze indicator with 40% and Ichimoku with 40%.

In Table 5, the results of the stock market price prediction by proposed methods based on the testing set (30% of the second part of the data) is shown, which is measured by different evaluation criteria. Based on the results, the Ridge and Linear Regression methods with all the R2, MSE, RMSE, MAE and MAPE evaluation criteria have the best stock market price prediction results. Also, the MLP Regression with Sequential Forward Selection and the MSE evaluation criteria, had the best performance.

In terms of improving stock price forecasts with regression methods along with Sequential Forward Selection and Sequential Backward Selection; MLPSF with MSE criteria improved by 56.47% compared to MLP regression with all indicators. Also, SVRSF with MSE criteria improved by 67.42 % and SVRSB with MSE criteria improved by 38.35% compared to SVR with all indicators.

LRSF with MSE improved by 76.9% and LRSB improved by 67.67% compared to Linear Regression with all indicators. RISF and RISB with MSE improved by 72.82% compared to Ridge regression with all the indicators. DTRSF with MSE improved by 83.17% and DTRSB with MSE improved by 23.24% compared to the DTR with all the indicators. KNNSF and KNNSB with MSE improved by 15.52% compared to KNN regression with all indicators. RFSF with MSE improved by 4% and RFSB with MSE improved by 6% compared to RF regression with all indicators. GBRSF also improved by 7% and GBSB by 2% compared to GBR with all indicators. Finally, ADASF and ADASB also had a 4% improvement over the ADA regression with all indicators. The results show that the use of the technical indicators selected in this study performs well for the stock market forecast, and the researchers can use these indicators instead of using all the indicators that require a lot of calculations to predict high -precision stock market prices. Now that the types of technical indicators have been identified by different regression methods to predict the stock market price, these technical indicators can be used with each machine learning method for better and higher precision predictions.

**Table 5** Stock market prediction by proposed methods on testing set with MAE, MSE, RMSE, MAPE and R2

| Model/ Metric | MSE | RMSE | MAE | MAPE | R2 |
|---|---|---|---|---|---|
| **LR** | 0/00105 | 0/0324 | 0/02338 | 0/03157 | 0/95252 |
| **LRSF** | 0/00025 | 0/01568 | 0/01636 | 0/03279 | 0/98991 |
| **LRSB** | 0/00034 | 0/01849 | 0/01452 | 0/019 | 0/98991 |
| **Lasso** | 0/38799 | 0/62289 | 0/60487 | 0/82337 | -16/5436 |
| **LOSF** | 0/38799 | 0/62289 | 0/60487 | 0/82337 | -16/5436 |
| **LOSB** | 0/38799 | 0/62289 | 0/60487 | 0/82337 | -16/5436 |
| **Ridge** | 0/00092 | 0/03038 | 0/02193 | 0/02988 | 0/95827 |
| **RISF** | 0/00025 | 0/01572 | 0/01197 | 0/02053 | 0/98991 |
| **RISB** | 0/00025 | 0/01583 | 0/01478 | 0/0236 | 0/98904 |
| **DTR** | 0/18562 | 0/43084 | 0/40069 | 0/52294 | -7/39338 |
| **DTRSF** | 0/15251 | 0/39052 | 0/3476 | 0/48819 | -6/14302 |
| **DTRSB** | 0/14247 | 0/37745 | 0/35791 | 0/46613 | -5/87599 |
| **KNN** | 0/17019 | 0/41254 | 0/38277 | 0/49943 | -6/69553 |
| **KNNSF** | 0/14376 | 0/37916 | 0/35005 | 0/64836 | -5/3447 |
| **KNNSB** | 0/14111 | 0/37565 | 0/34622 | 0/70377 | -5/50056 |
| **MLP** | 0/01475 | 0/12144 | 0/11035 | 0/14384 | 0/33313 |
| **MLPSF** | 0/00642 | 0/28459 | 0/12435 | 0/27462 | 0/6091 |
| **MLPSB** | 0/04832 | 0/21981 | 0/08808 | 0/13822 | 0/68399 |
| **SVR** | 0/31864 | 0/56448 | 0/54128 | 0/7261 | -13/4078 |
| **SVRSF** | 0/1038 | 0/32217 | 0/33132 | 0/50467 | -4/60466 |
| **SVRSB** | 0/19642 | 0/4432 | 0/34233 | 0/52814 | -4/76679 |
| **RFR** | 0/15987 | 0/39983 | 0/37191 | 0/48577 | -6/22875 |
| **RFRSF** | 0/15346 | 0/38645 | 0/34745 | 0/50173 | -5/53421 |
| **RFRSB** | 0/15015 | 0/3875 | 0/37797 | 0/4695 | -5/65753 |
| **GBR** | 0/18272 | 0/42746 | 0/40122 | 0/52813 | -7/26228 |
| **GBRSF** | 0/17002 | 0/45841 | 0/38735 | 0/66709 | -6/80044 |
| **GBRSB** | 0/17881 | 0/42285 | 0/40297 | 0/52347 | -6/77006 |
| **ADA** | 0/15983 | 0/39979 | 0/37341 | 0/49021 | -6/22701 |
| **ADASF** | 0/15403 | 0/39246 | 0/36379 | 0/64818 | -5/93106 |
| **ADASB** | 0/15269 | 0/39075 | 0/36441 | 0/47458 | -5/91784 |

In Figure 8, the best performance for MLPSF and LRSF with MSE error-based evaluation criteria is shown. So, adding past technical indicators as features can improve the performance of the model.

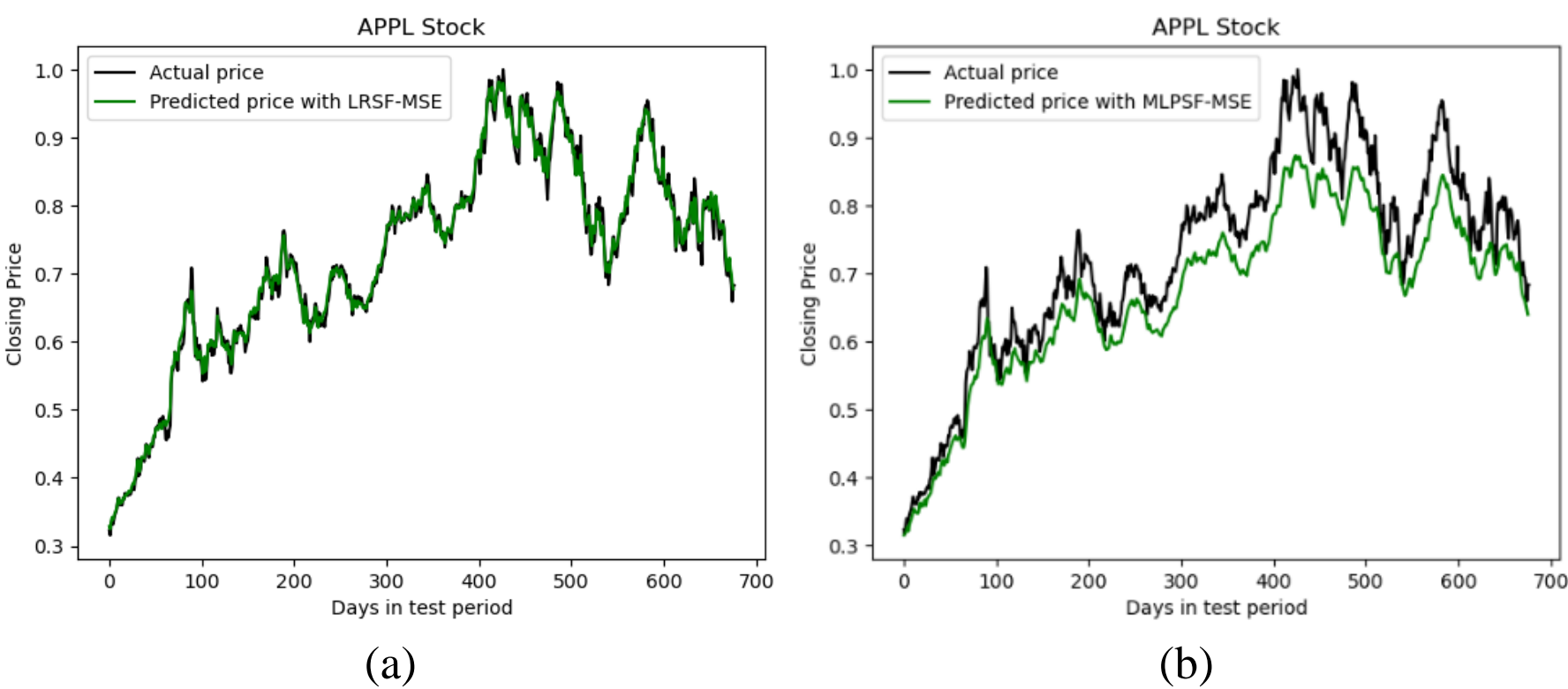

**Fig. 8** (a) best performance for LRSF and (b) best performance for MLPSF with MSE evaluation criteria

Best combination of technical indicator for MLPSF are: Kdj, bbands, vortex, wcp, vwap, qqe, decay, mcgd, supertrend, uo, ssf, pvt, rvgi, hwc, fisher, aobv, cfo, willr, Sine wma, linreg, efi, rma.

Best combination of technical indicator for LRSF are: squeeze_pro, aobv, pvo, pvt, rvgi, kama, decay, mom, ichimoku, bbands, cfo, trima, hilo, dpo, ebsw, wcp, thermo, stc, pvr, psar, natr, midpoint, pwma.

After understanding the proper composition of the indicators, with MLPSF, SVRSF and LRSF methods, we will try to find the best time window to predict the closing price. Here the size window size is 1, 3, 7,15 and 30 days (Figure 9). The best window size is w=3. Results comparison with existing work is shown in Table 5. Based on Table 6, proposed method has least error with MSE criteria.

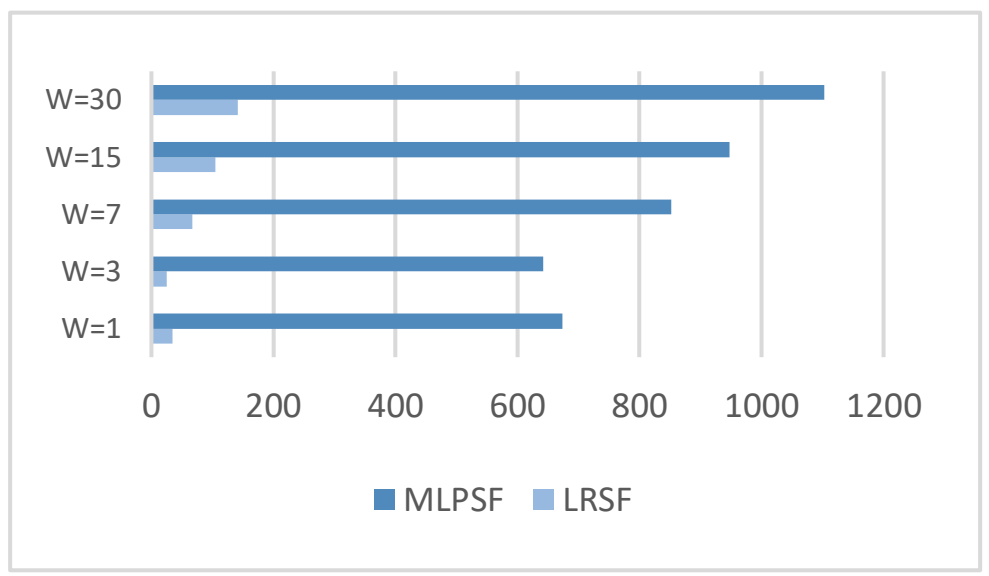

**Fig. 9** Comparison of time window size with MSE criteria (MSE*$10^5$)

**Table 6** Comparison of results with Existing work

| Methods | MSE | MAE |
|---|---|---|
| **C-E-SVR&RF [23]** | 0/1743 | 0/3223 |
| **E-SVR&RF [23]** | 0/1840 | 0/3269 |
| **SVRSF** | 0/1038 | 0/33132 |
| **RFRSF** | 0/15346 | 0/34745 |
| **ANN [42]** | - | 15/1221 |
| **MLPSF** | 0/00642 | 0/12435 |

### 4.1 The empirical result of the feature selection

We selected the optimal feature subset for each method based on the relationship between regression model performance and the dataset. With the selection of a 3-day time window, the number of features increased significantly, which led to the creation of additional features, thus substantially increasing the computational workload. The application of Forward and backward feature selection methods also demanded substantial computational resources, and the Sequential Backwards Selection method slowed down the process.

By applying both Sequential Forwards Selection and Backward Selection, as shown in the results in Table 4 (in the appendix), the number of features was reduced for each model according to the evaluation criterion, while preserving the features critical to the model. The selected features were more important than those deleted, ensuring that the best subset of features was always chosen from the most important ones.

Secondly, the selected features were used as the input to predict stock prices with different models, and the performance of each model was evaluated based on error-based evaluation criteria. The results showed an improvement in model performance compared to the original model that used all features. The use of technical indicators, as listed in Table 4 (in the appendix), along with different models and the 3-day time window, is recommended for future applications.

## 5 Conclusion

The results of applying Forward and Backward Feature selection to 123 technical indicators on Apple's stock data demonstrated significant performance improvements across various regression methods. The improvements observed are as follows (based on Table 5); MLPSF outperformed MLP by 56.47%, SVRSF outperformed SVR by 67.42%, LRSF outperformed Linear Regression by 76.7%, RISF outperformed Ridge Regression by 72.82%, DTRSB outperformed DTR by 24.23%, KNNSB outperformed KNN by 15.52%, RFRSB outperformed RF by 6%, GBRSF outperformed GBR by 7%, ADASF and ADASB outperformed ADA by 4%. Additionally, Ridge Regression and Linear Regression achieved the best performance with an MSE of 0.00025.

The Sequential Backward Selection (SBS) algorithm was observed to be slower than Sequential Forward Selection (SFS) due to its higher memory demands at the initial stage. Various machine learning models exhibited different feature selection preferences, with each model selecting features based on its respective evaluation criteria.

Several technical indicators were consistently selected across different regression models, including: Squeeze_pro, Percentage Price Oscillator, Thermo, Decay, Archer On-Balance Volume, Bollinger Bands, Squeeze, and Ichimoku.

MLP Regression combined with Sequential Forward Selection (SFS) and MSE evaluation yielded the best performance, with an MSE of 0.00642 (as shown in Table 5). Similarly, SVR with SFS and the MSE (SVRSF) evaluation criterion demonstrated significant improvements compared to the results obtained using all technical indicators.

Based on the findings of this study, it is recommended to use feature selection techniques (especially SFS) along with models such as MLP, SVR, Ridge, and Linear Regression in combination with the identified technical indicators to enhance the accuracy and reliability of stock market predictions.

For future research, it would be valuable to explore the study of different time periods and financial assets, as well as to examine the features selected using different time windows to predict various regression models. It is also recommended to focus on the indicators selected in this study (especially those that perform well with multilayer perceptron regression) for predictions using various neural networks. Additionally, by analyzing the selected features across different methods, the most important features can be identified, and this will be explored further in future work.

## Appendixes

**TABLE 1** The definition of the 123 indicators in the data sets

| No. | Indicator | Name | No. | Indicator | Name |
|---|---|---|---|---|---|
| 1 | Aberration | aberration | 63 | Normalized Average True Range | natr |
| 2 | Absolute Price Oscillator | apo | 64 | Ohlc4 | ohlc4 |
| 3 | Acceleration Bands | accbands | 65 | On-Balance Volume | obv |
| 4 | Accumulation/Distribution Index | ad | 66 | Parabolic Stop and Reverse | psar |
| 5 | Accumulation/Distribution Oscillator | adosc | 67 | Pascal's Weighted Moving Average (PWMA) | pwma |
| 6 | ALMA | alma | 68 | Percentage Price Oscillator | ppo |
| 7 | Archer On-Balance Volume | aobv | 69 | Percentage Volume Oscillator | pvo |
| 8 | Aroon | aroon | 70 | Positive Volume Index | pvi |
| 9 | Average Directional Movement Index | adx | 71 | Pretty Good Oscillator | pgo |
| 10 | Average True Range | atr | 72 | Price Distance | pdist |
| 11 | Awesome Oscillator | ao | 73 | Price Volume Rank | pvr |
| 12 | Balance of Power | bop | 74 | Price Volume Trend | pvt |
| 13 | Bias | bias | 75 | Price-Volume | pvol |
| 14 | Bollinger Bands | bbands | 76 | Psychological Line | psl |
| 15 | BRAR | brar | 77 | Q Stick | qstick |
| 16 | Center of Gravity | cg | 78 | Quantitative Qualitative Estimation | qqe |
| 17 | Chaikin Money Flow | cmf | 79 | Rate of Change | roc |
| 18 | Chande Forecast Oscillator | cfo | 80 | Relative Strength Index | rsi |
| 19 | Chande Kroll Stop | cksp | 81 | Relative Strength Xtra | rsx |
| 20 | Chande Momentum Oscillator | cmo | 82 | Relative Vigor Index | rvgi |
| 21 | Choppiness Index | chop | 83 | Relative Volatility Index | rvi |
| 22 | Commodity Channel Index | cci | 84 | RMA | rma |
| 23 | Coppock Curve | coppock | 85 | Rollin_Z_Score | zscore |
| 24 | Correlation Trend Indicator | cti | 86 | Rolling Standard Deviation | stdev |
| 25 | Decay | decay | 87 | Rolling_Kurtosis | kurtosis |
| 26 | Decreasing | decreasing | 88 | Rolling_Mean_Absolute_Deviation | mad |
| 27 | DEMA | dema | 89 | Rolling_Median | median |
| 28 | Detrended Price Oscillator | dpo | 90 | Rolling_Quantile | quantile |
| 29 | Directional Movement | dm | 91 | Rolling_Skew | skew |
| 30 | Donchian Channel | donchian | 92 | Rolling_Variance | variance |
| 31 | Ease of Movement | eom | 93 | Schaff Trend Cycle | stc |
| 32 | EBSW | ebsw | 94 | Simple Moving Average (SMA) | sma |
| 33 | Efficiency Ratio | er | 95 | Slope | slope |

| 34 | Ehler's Super Smoother Filter (SSF) | ssf | 96 | SMI Ergodic | smi |
|---|---|---|---|---|---|
| 35 | Elder Ray Index | eri | 97 | Squeeze | squeeze |
| 36 | Elder's Force Index | efi | 98 | Squeeze Pro | squeeze_pro |
| 37 | EMA | ema | 99 | Stochastic Oscillator | stoch |
| 38 | Fibonacci's Weighted Moving Average | fwma | 100 | Stochastic RSI | stochrsi |
| 39 | Fisher Transform | fisher | 101 | Supertrend | supertrend |
| 40 | Gann HiLo Activator | hilo | 102 | Sine Weighted Moving Average (SWMA) | Sine wma |
| 41 | Hl2 Indicator | hl2 | 103 | Symmetric Weighted Moving Average (SWMA) | swma |
| 42 | Hlc3 Indicator | hlc3 | 104 | T3 | t3 |
| 43 | Holt-Winter Channel | hwc | 105 | Td_seq | td_seq |
| 44 | Hull Moving Average | hma | 106 | TEMA | tema |
| 45 | HWMA) Holt-Winter Moving Average( | hwma | 107 | Thermo | thermo |
| 46 | Ichimoku | ichimoku | 108 | TRIMA | trima |
| 47 | Increasing | increasing | 109 | Trix | trix |
| 48 | Indicator :Archer Moving Averages Trends | amat | 110 | True Range | true_range |
| 49 | Kaufman's Adaptive Moving Average | kama | 111 | True strength index | tsi |
| 50 | Kdj | kdj | 112 | TTM Trend | ttm_trend |
| 51 | Keltner Channel | kc | 113 | Ulcer Index | ui |
| 52 | Klinger Volume Oscillator | kvo | 114 | Ultimate Oscillator | uo |
| 53 | KST Oscillator | kst | 115 | Vertical Horizontal Filter | vhf |
| 54 | Linear Regression Moving Average | linreg | 116 | VIDYA | vidya |
| 55 | Mass Index | massi | 117 | Volume Profile | vp |
| 56 | McGinley Dynamic Indicator | mcgd | 118 | Volume Weighted Average Price (VWAP) | vwap |
| 57 | Midpoint | midpoint | 119 | Vortex | vortex |
| 58 | Midprice | midprice | 120 | Weighted Closing Price (WCP) | wcp |
| 59 | Momentum | mom | 121 | Williams% R | willr |
| 60 | Money Flow Index | mfi | 122 | WMA | wma |
| 61 | Moving Average Convergence Divergence | macd | 123 | Zero lag exponential moving average | zlma |
| 62 | Negative Volume Index | nvi | | | |

**TABLE 4** The selected features by Sequential Forward Selection and Sequential Backwards Selection on Apple's data set

| Model | Evaluation Criteria | Selected Features | Model | Evaluation Criteria | Selected Features |
|---|---|---|---|---|---|
| | **MSE** | 98, 7, 69, 74, 82, 49, 25, 59, 46, 14, 18, 108, 40, 28, 32, 120, 107, 93, 73, 66, 63, 57, 67 | | **MSE** | 2, 96, 99, 101, 61, 123, 114, 104, 95, 31, 50, 3, 20, 100, 78, 57, 53, 6, 99, 109, 45, 28 |
| | **RMSE** | 98, 7, 69, 74, 82, 49, 25, 59, 46, 14, 18, 108, 40, 28, 32, 120, 107, 93, 73, 66, 63, 57, 67 | | **RMSE** | 2, 96, 99, 101, 61, 123, 114, 104, 95, 31, 50, 3, 20, 100, 78, 57, 53, 6, 109, 45, 28 |
| **LRSF** | **MAE** | 98, 7, 121, 93, 60, 53, 35, 46, 50, 14, 18, 17, 120, 75, 62, 28, 97, 32, 101, 107, 86, 57, 88 | **LRSB** | **MAE** | 3, 4, 92, 96, 101, 116, 107, 86, 43, 35, 9, 67, 111, 95, 76, 74, 81, 79, 66, 62, 58, 44, 36, 28, 25, 56, 46 |
| | **MAPE** | 98, 97, 50, 117, 105, 14, 96, 109, 101, 66 | | **MAPE** | 18, 92, 111, 109, 107, 69, 102, 82, 28, 67, 50, 7, 37, 99, 80, 61, 116, 89, 53, 68, 98, 4, 24, 86, 81, 39, 36, 9, 31 |
| | **R2** | 17, 121, 110, 93, 91, 72, 45, 35, 28, 14, 61, 95, 75, 78, 58, 41, 2, 32, 100, 69, 120, 82, 57, 25 | | **R2** | 17, 121, 110, 93, 91, 72, 45, 35, 28, 14, 61, 95, 75, 78, 58, 41, 2, 32, 100, 69, 120, 82, 57, 25 |
| | **MSE** | 98, 7, 14, 121, 75, 78, 66, 42, 46, 107, 73, 57, 52, 40, 28, 25, 8, 32, 119, 120, 35 | | **MSE** | 11, 9, 12, 16, 119, 120, 78, 52, 53, 43, 28, 84, 1, 5, 2, 6, 18, 24, 92, 96, 99, 95, 74, 66, 36 |
| | **RMSE** | 98, 7, 14, 121, 75, 78, 66, 42, 46, 107, 73, 57, 52, 40, 28, 25, 8, 32, 119, 120, 35 | | **RMSE** | 11, 9, 12, 16, 119, 120, 78, 52, 53, 43, 28, 84, 1, 5, 2, 6, 18, 24, 92, 96, 99, 95, 74, 66, 36 |
| **RISF** | **MAE** | 98, 77, 14, 121, 99, 101, 93, 69, 78, 66, 72, 39, 35, 8, 107, 82, 36, 28, 32, 120, 118, 112, 86, 52 | **RISB** | **MAE** | 3, 18, 19, 17, 101, 109, 115, 104, 64, 28, 31, 67, 2, 111, 51, 35, 25, 7, 4, 24, 85, 61, 95, 70, 62, 55, 36 |
| | **MAPE** | 98, 32, 122, 120, 107, 105, 69, 74, 78, 90, 66, 82, 57, 19, 75, 99, 109, 44, 28, 46 | | **MAPE** | 1, 4, 5, 10, 12, 119, 121, 80, 78, 81, 40, 68, 19, 20, 76, 55, 45, 9, 111, 101, 61, 104, 66, 51, 28, 46 |
| | **R2** | 17, 121, 110, 93, 91, 72, 45, 35, 28, 14, 61, 95, 75, 78, 58, 41, 2, 32, 100, 69, 120, 82, 57, 25 | | **R2** | 13, 114, 78, 53, 28, 25, 5, 2, 20, 99, 101, 70, 62, 60, 56, 65, 7, 4, 6, 19, 24, 61, 58, 52, 49 |
| | **MSE** | 68, 98, 77, 1, 97, 50, 11, 65, 7, 3 | | **MSE** | 68, 98, 77, 1, 97, 50, 11, 65, 7, 3 |
| | **RMSE** | 68, 98, 77, 1, 97, 50, 11, 65, 7, 3 | | **RMSE** | 68, 98, 77, 1, 97, 50, 11, 65, 7, 3 |
| **LOSF** | **MAE** | 68, 98, 77, 1, 97, 50, 11, 65, 7, 3 | **LOSB** | **MAE** | 68, 98, 77, 1, 97, 50, 11, 65, 7, 3 |
| | **MAPE** | 68, 98, 77, 1, 97, 50, 11, 65, 7, 3 | | **MAPE** | 68, 98, 77, 1, 97, 50, 11, 65, 7, 3 |
| | **R2** | 68, 98, 77, 1, 97, 50, 11, 65, 7, 3 | | **R2** | 68, 4, 32, 103, 69, 44, 39, 26, 46, 6, 76, 74, 58, 57, 54, 52, 42, 41, 7, 3, 106, 123, 120, 115, 112, 105, 91, 43, 33 |
| | **MSE** | 68, 98, 97, 7, 20, 85, 101, 120, 117, 107, 95, 66, 72, 57, 28, 30, 106, 112, 105, 26, 25 | | **MSE** | 98, 113, 107, 75, 82, 46, 1, 16, 37, 99, 34, 53, 47, 35, 4, 10, 19, 76, 106, 43, 42, 39, 25, 88 |
| | **RMSE** | 68, 98, 97, 7, 20, 85, 101, 120, 117, 107, 95, 66, 72, 57, 28, 30, 106, 112, 105, 26, 25 | | **RMSE** | 98, 113, 107, 75, 82, 46, 1, 16, 37, 99, 34, 53, 47, 35, 4, 10, 19, 76, 106, 43, 42, 39, 25, 88 |
| **DTRSF** | **MAE** | 98, 119, 92, 85, 109, 107, 73, 52, 40, 5, 101, 86, 93, 75, 83, 90, 71, 30, 80, 105, 79, 63, 55, 41, 28, 25, 46 | **DTRSB** | **MAE** | 98, 1, 97, 7, 20, 27, 100, 61, 110, 79, 66, 84, 59, 50, 6, 19, 96, 81, 55, 9, 13, 58, 57, 42, 25, 31 |
| | **MAPE** | 97, 14, 27, 30, 117, 110, 107, 105, 86, 78, 44, 40, 28, 46, 73, 68, 7, 101, 83, 43, 25 | | **MAPE** | 2, 13, 21, 23, 85, 96, 101, 55, 53, 44, 68, 50, 7, 14, 123, 120, 28, 94, 46, 1, 99, 103, 117, 87, 42, 25 |
| | **R2** | 9, 19, 30, 119, 92, 115, 107, 78, 102, 44, 33, 46, 4, 111, 101, 80, 109, 91, 90, 88, 11, 7, 93, 66, 71, 89, 41, 25 | | **R2** | 68, 98, 65, 3, 20, 111, 95, 75, 45, 50, 6, 101, 69, 116, 1, 4, 17, 30, 118, 90, 66, 58, 46 |
| | **MSE** | 98, 97, 16, 117, 105, 106, 120, 25 | | **MSE** | 107, 14, 106, 123, 120, 118, 42, 25, 1, 27, 109, 74, 64, 58, 57, 54, 43, 44, 41, 38 |
| | **RMSE** | 98, 97, 16, 117, 105, 106, 120, 25 | | **RMSE** | 107, 14, 106, 123, 120, 118, 42, 25, 1, 27, 109, 74, 64, 58, 57, 54, 43, 44, 41, 38 |
| **KNNSF** | **MAE** | 98, 97, 117, 105, 34, 25 | **KNNSB** | **MAE** | 1, 10, 14, 107, 4, 123, 120, 57, 44, 25, 27, 37, 103, 106, 118, 45, 43, 64, 54, 42, 41, 38 |
| | **MAPE** | 98, 50, 101, 117, 107, 105, 78, 79, 66, 40, 31, 96, 35 | | **MAPE** | 98, 8, 101, 107, 78, 66, 40, 97, 7, 117 |
| | **R2** | 98, 97, 117, 105, 34, 25 | | **R2** | 123, 69, 72, 28, 14, 106, 120, 118, 107, 42, 25, 1, 4, 27, 103, 122, 64, 58, 57, 54, 44, 38, 67 |
| | **MSE** | 50, 14, 119, 120, 118, 78, 25, 56, 101, 114, 34, 74, 82, 43, 39, 7, 18, 121, 102, 54, 36, 84 | | **MSE** | 68, 7, 2, 13, 21, 32, 96, 113, 78, 53, 88, 14, 23, 30, 99, 117, 9, 115 |
| | **RMSE** | 68, 12, 17, 23, 96, 110, 78, 53, 100, 86, 95, 35, 46, 1, 50, 65, 13, 16, 113, 107, 66, 40, 28, 9, 59 | | **RMSE** | 68, 7, 2, 13, 21, 32, 96, 113, 78, 53, 88, 14, 23, 30, 99, 117, 9, 115 |
| **MLPSF** | **MAE** | 14, 12, 120, 43, 35, 33, 25, 46, 115, 95, 72, 42, 9, 26, 68, 4, 118, 45, 83, 62, 57, 49, 41, 56 | **MLPSB** | **MAE** | 32, 37, 85, 118, 117, 105, 45, 43, 42, 38, 56, 46, 50, 96, 64, 62, 54, 25, 94 |
| | **MAPE** | 7, 8, 18, 107, 69, 79, 66, 72, 53, 49, 35, 68, 98, 97, 14, 16, 82, 52, 47, 13, 96, 101, 81 | | **MAPE** | 68, 2, 122, 113, 78, 53, 45, 23, 123, 117, 114, 110, 91, 51, 13, 61, 89, 52 |

| | R2 | 1, 101, 64, 45, 43, 44, 46, 9, 91, 25, 56, 14, 118, 116, 69, 62, 58, 47, 42, 41, 59, 67 | | R2 | 16, 101, 120, 118, 117, 105, 64, 45, 43, 42, 46, 110, 107, 33, 9, 41, 56 |
|---|---|---|---|---|---|
| | **MSE** | 68, 98, 117, 105, 46, 107, 34, 66, 56 | | **MSE** | 1, 14, 58, 57, 27, 106, 122, 123, 38, 46, 37, 103, 104, 94 |
| | **RMSE** | 68, 98, 117, 105, 46, 107, 34, 66, 56 | | **RMSE** | 1, 14, 58, 57, 27, 106, 122, 123, 46, 37, 103, 104, 38, 94 |
| **SVRSF** | **MAE** | 98, 117, 105, 72, 62, 46, 32 | **SVRSB** | **MAE** | 1, 10, 12, 86, 71, 9, 18, 107, 53, 35, 59, 98, 77, 97, 101, 69, 82, 54, 33 |
| | **MAPE** | 98, 117, 107, 105, 4, 14, 90 | | **MAPE** | 8, 32, 109, 107, 73, 66, 52, 40, 9, 12, 85, 63, 98, 14, 121, 61, 75 |
| | **R2** | 98, 117, 105, 72, 62, 46, 32 | | **R2** | 98, 77, 1, 97, 10, 12, 121, 92, 85, 9, 39, 106, 123, 82, 71, 44, 59 |
| | **MSE** | 98, 110, 105, 72, 1, 97, 9, 14, 12, 111, 117, 89, 46, 68, 23, 120, 118, 107, 69, 57, 43, 40, 25 | | **MSE** | 3, 5, 8, 108, 107, 69, 57, 77, 1, 65, 12, 34, 78, 62, 33, 19, 80, 122, 64, 43, 41, 40, 26, 59 |
| | **RMSE** | 98, 110, 105, 72, 1, 97, 9, 14, 12, 111, 117, 89, 46, 68, 23, 120, 118, 107, 69, 57, 43, 40, 25 | | **RMSE** | 3, 5, 8, 108, 107, 69, 57, 77, 1, 65, 12, 34, 78, 62, 33, 19, 80, 122, 64, 43, 41, 40, 26, 59 |
| **RFRSF** | **MAE** | 98, 97, 101, 110, 107, 26, 46, 117, 105, 83, 7, 120, 118, 66, 58, 25 | **RFRSB** | **MAE** | 68, 8, 14, 13, 99, 79, 62, 40, 39, 77, 97, 50, 7, 96, 123, 120, 34, 53, 101, 122, 115, 87, 35 |
| | **MAPE** | 22, 119, 117, 114, 76, 60, 39, 46, 98, 77, 97, 65, 3, 23, 123, 107, 52, 53, 45, 26, 88, 4, 30, 109, 113, 9 | | **MAPE** | 68, 65, 43, 9, 84, 6, 99, 101, 117, 116, 107, 93, 95, 79, 66, 57, 36, 28, 25, 4, 2, 16, 27, 96, 118, 34, 74, 56 |
| | **R2** | 97, 8, 117, 47, 31, 98, 107, 26, 68, 4, 123, 120, 118, 66, 57, 25 | | **R2** | 10, 111, 101, 115, 110, 38, 31, 46, 68, 98, 3, 14, 61, 117, 67, 1, 9, 106, 120, 105, 64, 63, 42, 41, 25 |
| | **MSE** | 97, 5, 14, 96, 105, 66, 62, 52, 9, 68, 18, 21, 107, 93, 64, 98, 76, 77, 7, 85, 100, 120, 78 | | **MSE** | 68, 7, 13, 30, 119, 93, 82, 62, 39, 98, 14, 61, 110, 74, 53, 59, 67, 88, 46, 97, 8, 18, 121, 92, 96, 69, 72, 40 |
| | **RMSE** | 68, 5, 12, 101, 61, 117, 93, 95, 76, 69, 78, 81, 66, 72, 59, 97, 9, 16, 121, 8, 24, 86, 87, 53 | | **RMSE** | 68, 7, 13, 30, 119, 93, 82, 62, 39, 98, 14, 61, 110, 74, 53, 59, 67, 88, 46, 97, 8, 18, 121, 92, 96, 69, 72, 40 |
| **GBRSF** | **MAE** | 68, 98, 77, 97, 50, 8, 17, 119, 117, 28, 13, 21, 24, 105, 91, 62, 7, 76, 25 | **GBRSB** | **MAE** | 68, 3, 92, 61, 53, 35, 28, 59, 97, 14, 13, 30, 93, 78, 82, 87, 39, 84, 94, 2, 111, 115, 66, 47, 40, 88, 46 |
| | **MAPE** | 98, 97, 8, 117, 112, 107, 105, 95, 78, 66, 47, 26, 7, 115, 40 | | **MAPE** | 68, 98, 77, 4, 6, 10, 19, 92, 96, 122, 114, 62, 53, 43, 40, 39, 59, 46, 119, 106, 118, 70, 74, 78, 25 |
| | **R2** | 68, 98, 97, 50, 7, 5, 9, 2, 8, 14, 111, 4, 24, 61, 93, 28, 31, 78, 25 | | **R2** | 1, 97, 96, 101, 113, 107, 46, 68, 122, 86, 62, 51, 40, 3, 4, 14, 106, 61, 123, 74, 57, 43, 42, 41, 39, 25 |
| | **MSE** | 68, 98, 8, 19, 61, 115, 69, 74, 83, 82, 41, 16, 109, 120, 71, 51, 45, 33, 25, 92, 96, 113, 86, 78, 52, 53 | | **MSE** | 68, 8, 14, 61, 107, 45, 36, 7, 4, 16, 92, 75, 79, 71, 87, 44, 40, 35, 46, 34, 90, 57, 89, 53 |
| | **RMSE** | 68, 98, 8, 19, 61, 115, 69, 74, 83, 82, 41, 16, 109, 120, 71, 51, 45, 33, 25, 92, 96, 113, 86, 78, 52, 53 | | **RMSE** | 68, 8, 14, 61, 107, 45, 36, 7, 4, 16, 92, 75, 79, 71, 87, 44, 40, 35, 46, 34, 90, 57, 89, 53 |
| **ADASF** | **MAE** | 68, 8, 123, 107, 93, 73, 78, 38, 28, 1, 97, 14, 23, 119, 66, 43, 31, 7, 103, 57, 54, 45 | **ADASB** | **MAE** | 50, 7, 14, 111, 100, 117, 105, 73, 78, 66, 72, 62, 1, 11, 22, 107, 75, 45, 39, 16, 24, 119, 120, 69, 93, 63, 33 |
| | **MAPE** | 98, 97, 117, 107, 105, 47, 40, 26 | | **MAPE** | 8, 99, 117, 110, 75, 73, 66, 88, 68, 13, 18, 96, 82, 69, 52, 43, 40, 26, 22, 20, 121, 73, 42, 35 |
| | **R2** | 98, 1, 21, 27, 107, 93, 76, 74, 44, 39, 14, 119, 45, 40, 25, 31, 59, 65, 9, 122, 117, 78, 57, 33, 46 | | **R2** | 6, 8, 14, 12, 16, 20, 99, 100, 93, 69, 71, 55, 35, 68, 50, 13, 117, 115, 62, 7, 85, 74, 25 |